# A Comparative Study of Correlation and Relativistic Effects on Atomic Ionization Energy

Mohamed Kahil[1,*], Fatima Fakih[1,**], Nabil Joudieh[1,2,3,***], and Nidal Chamoun[4,5,****]

[1]Department of Physics, Faculty of Sciences, Damascus University, Damascus, Syria
[2]Laser Spectroscopy Laboratory, Higher Institute for Laser Research and Applications, Damascus University, Damascus, Syria
[3]Faculty of Pharmacy, International University for Science and Technology, Damascus, Syria
[4]Department of Statistics, Faculty of Sciences, Damascus University, Damascus, Syria
[5]CASP, Antioch Syrian University, Maaret Saidnaya, Damascus, Syria

*Emails:(\*) mohamed.kahel@damascusuniversity.edu.sy, (\*\*) fatima.fakih.du@gmail.com, (\*\*\*) njoudieh@yahoo.fr,*

*(Corresponding author \*\*\*\*) chamoun@uni-bonn.de*

## Abstract

This study investigates the interplay between relativistic effects and electron correlation effects on the first ionization energies of heavy atoms (Au through Rn, Z = 79–86). We perform two complementary analyses: (1) comparing relativistic corrections computed at both the Hartree-Fock (HF) and coupled cluster CCSD(T) levels to assess how correlation influences the magnitude of relativistic corrections, and (2) comparing correlation corrections computed within both non-relativistic and relativistic frameworks to determine how relativity influences the magnitude of correlation corrections.

Our results reveal a striking non-linear relationship between these two effects. Specifically, the combined effect of relativity and correlation on ionization energy does not equal the sum of their individual contributions. This non-additivity indicates that relativistic and correlation effects are not independent; they interact in complex ways that depend on the atomic system. We find that for some atoms, the two effects enhance each other, while for others they partially cancel. Moreover, the order in which one may add "separate" effects also counts, in that adding "pure" relativistic effects to the remaining outcome (including correlation) would give a different result than when adding "pure" correlation effects to the remaining outcome (including relativity).

These findings demonstrate that relativistic and correlation effects are inherently non-additive, reflecting the non-linearity of the quantum many-body problem. Accurate computational predictions of ionization energies in heavy-element systems thus require simultaneous treatment of both effects rather than treating them as independent contributions

## 1. Introduction

The accurate prediction of atomic properties, particularly ionization energies, requires careful treatment of quantum mechanical effects. Two of the most important contributions are electron correlation and relativistic effects. The Hartree-Fock (HF) method, which may incorporate relativistic corrections important in the study of heavy-element systems, provides a mean-field description of electron-electron interactions but neglects the instantaneous correlation between electrons. Coupled cluster (CC) theory, particularly the CCSD(T) method ("coupled cluster with singles, doubles and perturbative triples"), is a well-established approach for including electron correlation effects [1].

The treatment of electron correlation and relativistic effects on equal footing is an active and important research area in computational electronic structure theory [2]. Previous studies have examined these effects separately or in combination for specific systems. For example, Schwerdtfeger and colleagues [3] demonstrated that relativistic effects on ionization energies exceeded correlation effects for gold atoms. Similarly, other work [4] showed that relativistic effects dominated correlation contributions for valence ionization energies in lead, although the reverse was true for the first ionization energy. The first systematic study comparing both relativistic and non-relativistic effects at the correlated level appeared in 1990 [5], followed by work in 1991 [6] showing that relativistic effects increase the ionization potential of Au by 1.708 eV while correlation effects increase it by 1.269 eV. These studies established that both effects are important, but they did not systematically investigate how these effects interact with each other.

Despite these previous studies, to our knowledge, no systematic investigation has examined the mutual interplay between relativistic and correlation effects—that is, how one effect influences the magnitude and behavior of the other. This is the central question we address in this work. We investigate three key aspects: (1) whether relativistic corrections to ionization energies differ when computed with correlated versus uncorrelated electrons, (2) whether correlation corrections differ when computed within relativistic versus non-relativistic frameworks, and (3) we compared the separate relativistic and correlations effects in what regards their corrections contrasted to experiment.

We focus on the heaviest main-group elements (Au through Rn, with atomic numbers Z = 79–86) because these elements exhibit chemical and physical properties distinctly different from their lighter counterparts, and relativistic effects are most pronounced in this region [7]. Our analysis employs the Dirac-Coulomb (DC) Hamiltonian for relativistic calculations and the Schrödinger Hamiltonian for non-relativistic calculations, combined with the CCSD(T) method for correlation effects, while HF methods neglects them. Our key finding is that relativistic and correlation effects are not independent: their combined effect is not additive, reflecting the inherent non-linearity of the quantum many-body problem. This conclusion aligns with earlier observations [8] that these contributions are not independent, and our work provides a systematic quantitative demonstration of this non-additivity across a series of heavy atoms.

Not only the sum of the two types of corrections is not equal to the correction when combining the two effects, but the order in which one adds the separate effects also counts, in that the addition of the "pure" correlation effects to the relativistic HF result getting a "total" correction, is different in general from the other way round by adding the "pure" relativistic corrections to the non-relativistic CCSD(T) result giving another "total" correction.

We acknowledge that earlier works have noted the non-independence of relativistic and correlation effects (see, e.g., Ref. [8]). The novelty of our work lies in the systematic, quantitative exploration of this non-additivity across the complete 6th-row main-group elements (Au–Rn) using the CCSD(T) method within both Dirac–Coulomb and non-relativistic frameworks, providing a comprehensive dataset and analysis that were previously lacking, especially for Bi–Rn

## 2. Computational Methods

All calculations were performed using the DIRAC 2025 program [9]. For relativistic calculations, we employed the Dirac-Coulomb (DC) Hamiltonian [10], while for non-relativistic calculations, we used the Schrödinger Hamiltonian (NR). Both sets of calculations utilized the dyall.4zp basis set [11] and a nuclear Gaussian charge distribution [12].

The electronic configurations of the studied atoms follow the pattern [Xe] $4f^{14}\ 5d^{10}\ 6s^x\ 6p^y$, where x and y vary depending on the element. For Au and Hg, x = 1 and 2 respectively, with y = 0. For elements from Tl through Rn, x = 2 and y ranges from 1 to 6. In the non-relativistic treatment, electrons occupy the three p orbitals with single occupancy followed by pairing. In the relativistic treatment, spin-orbit coupling causes the three p orbitals to split into one $p_{1/2}$ and two $p_{3/2}$ orbitals, with the $p_{1/2}$ orbitals being more stable and filled before the $p_{3/2}$ orbitals. The latter orbitals are occupied according to average-of-configuration (AOC) manner. For example the considered electron configuration for Bi atom in the relativistic and non-relativistic cases is as follows, where the highest electron in the relativistic case, to be put in the $p_{3/2}$ four spinors, is allocated the spinor occupation value of ¼ in DIRAC:

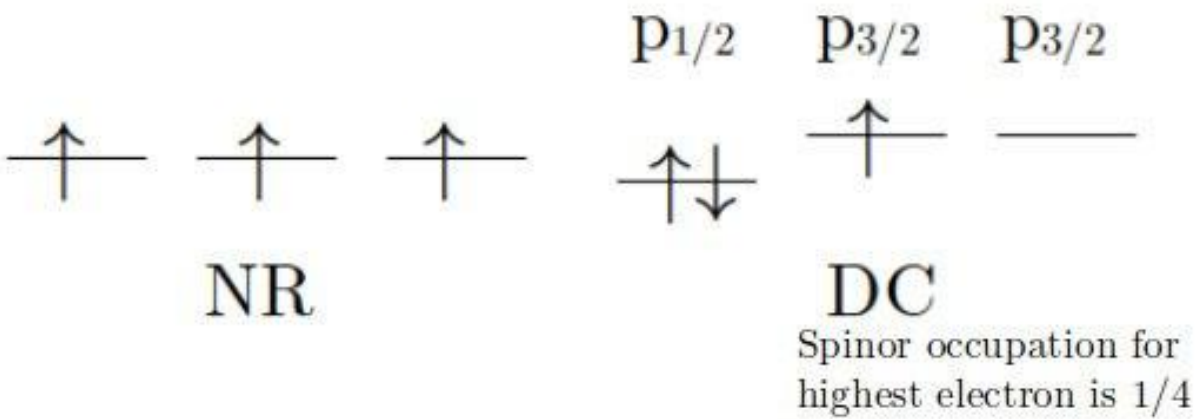

Schema 1: Set up of Spinor Occupation

To specify the active set of spinors in the relativistic calculations, we retained the default DIRAC setup of '-10.0 20.0 1.0', which selects all orbitals with energies between -10.0 and +20.0 Hartree (atomic units) with a minimum energy gap of 1.0 Hartree. This choice ensures inclusion of chemically relevant valence and sub-valence orbitals while excluding high-energy virtual orbitals that contribute negligibly. For the Au atom, this resulted in 87 virtual orbitals and 33 correlated electrons, corresponding to the electrons beyond the closed [Xe] core: the $4f^{14}$ shell (14 electrons), the $5d^{10}$ shell (10 electrons), and the valence $6s^1$ electron, in addition to 8 electrons already in the [Xe] core: $5s^2 5p^6$. The number of correlated electrons increases by one for each successive atom in the series. For consistency, the non-relativistic calculations were configured to include the same number of correlated electrons and virtual orbitals as their relativistic counterparts. Test calculations with different number of correlated electrons changed the results by around 0.1% confirming the robustness of our results.

Actually, electronic correlations become relatively pronounced in valence and neighbouring orbitals compared to core ones, due to many factors including:

- Distance & Energy: Valence electrons are in the highest energy levels, farthest from the nucleus, meaning they have higher potential energy and are less attracted to the core.
- Shielding & Effective Nuclear Charge: Core electrons shield the inner shells, so valence electrons experience a weaker effective nuclear pull, making their motions less constrained and more easily influenced by other electrons.

We have performed additional calculations to assess the basis-set convergence for Hg using the dyall.3zp (getting 10.33 eV) and dyall.4zp (10.34 eV) basis sets. The difference in ionization energy between these basis sets is ≤ 0.01 eV (∼0.1%), indicating that the dyall.4zp basis is near the basis-set limit for these properties.

In order to test the virtual space cutoff sensitivity, we found that the deviation due to changing the virtual orbitals number can be neglected with a corresponding contributing effect not exceeding 0.9%. Actually, the CCSD(T) method gave for Hg ionization potential with 82 virtual orbitals the value 10.34 eV, which remains the same, up to second significative digit, when the number of virtual orbitals was increased to 143.

We chose to investigate the ionization energy (IE) since its experimental values are available and determined to an acceptable accuracy. IE was calculated using the standard definition:

$$IE(A) = E(A^+) - E(A)$$

where $E(A^+)$ and $E(A)$ represent the total energies of the cation and neutral atom, respectively.

## 3. Results and Discussion

### 3.1 Relativistic Effects in Correlated and Uncorrelated Atomic Systems

Table 1 presents the calculated ionization energies for atoms Z = 79–86 computed using both the HF and CCSD(T) methods, with and without relativistic corrections, as well as the corresponding absolute and relative differences. The values in brackets represent the percentage deviation from experimental values [13].

A clear pattern emerges from the data. For elements with atomic numbers 79–82 (Au, Hg, Tl, Pb), relativistic effects consistently increase the ionization energy, whether computed at the HF or CCSD(T) level. This increase is attributed to relativistic contraction and stabilization of the valence s and $p_{1/2}$ orbitals [14]. The relativistic contraction of these orbitals makes electrons more tightly bound to the nucleus, requiring more energy to remove them. Conversely, for elements with atomic numbers 83–86 (Bi, Po, At, Rn), with the exception of Po in the presence of correlation, which have valence electrons in $p_{3/2}$ orbitals, relativistic effects decrease the ionization energy. This reversal reflects the different orbital energetics when spin-orbit coupling is included: the $p_{3/2}$ orbitals are less stable than the $p_{1/2}$ orbitals due to relativistic effects. Actually, the increase for Au–Pb is dominated by scalar relativistic contraction, which stabilizes s and $p_{1/2}$ orbitals, whereas the decrease for Bi–Rn is primarily due to spin-orbit coupling, which raises the energy of $p_{3/2}$ orbitals relative to their non-relativistic counterparts.

Regarding agreement with experimental values, we see that including relativistic corrections at both HF and CCSD(T) levels improves the predictions for atoms ending in s and $p_{1/2}$ orbitals (Au through Pb). However, for atoms ending in $p_{3/2}$ orbitals (Bi through Rn), the situation is more complex. At the HF level, relativistic corrections worsen the agreement with experiment, while at the CCSD(T) level, they generally improve it (with the exception of At).

Notably, for Bi, Po, At and Rn, the non-relativistic HF values show surprisingly good agreement with experiment, better than when relativistic and/or correlation effects are included. This unexpected result may suggest that error cancellation occurs in these cases.

Table 1 : Relativistic and non-relativistic values of ionization energies(eV) at the HF and CCSD(T) methods for $_z$X atoms (where Z is the atomic number ranging from 79(Au) to 86(Rn)). Values in brackets denote the discrepancy with experimental values[13]. Δ represents the corresponding relative change.

|  | $_{79}$Au | $_{80}$Hg | $_{81}$Tl | $_{82}$Pb | $_{83}$Bi | $_{84}$Po | $_{85}$At | $_{86}$Rn |
|---|---|---|---|---|---|---|---|---|
| HF-NR | 5.915 | 6.829 | 5.068 | 6.208 | 7.358 | 8.532 | 9.738 | 10.98 |
|  | (35.8%) | (34.5%) | (16.9%) | (16.2%) | (0.9%) | (1.4%) | (1.0%) | (2.2%) |
| HF-DC | 7.692 | 8.550 | 5.612 | 6.521 | 6.470 | 7.564 | 8.679 | 9.820 |
|  | (16.5%) | (18.0%) | (8.0%) | (12.0%) | (11.3%) | (10.1%) | (9.9%) | (8.6%) |
| (DC - NR)$_{HF}$ | 1.777 | 1.496 | 0.544 | 0.313 | -0.888 | -0.968 | -1.059 | -1.160 |
| Δ% | 30.04 | 21.91 | 10.73 | 5.04 | -12.07 | -11.35 | -10.87 | -10.56 |
| CC-NR | 7.019 | 8.326 | 5.404 | 6.016 | 8.660 | 8.084 | 9.758 | 11.47 |
|  | (23.8%) | (20.2%) | (11.4%) | (18.8%) | (18.7%) | (3.89%) | (1.2%) | (6.8%) |
| CC-DC | 9.156 | 10.34 | 5.926 | 7.226 | 7.066 | 8.147 | 9.027 | 10.45 |
|  | (0.7%) | (0.9%) | (2.9%) | (2.4%) | (3.1%) | (3.1%) | (6.4%) | (2.7%) |
| (DC -NR)$_{CC}$ | 2.137 | 2.015 | 0.522 | 1.210 | -1.594 | 0.063 | -0.731 | -1.02 |
| Δ% | 30.45 | 24.20 | 9.66 | 20.11 | -18.41 | 0.78 | -7.49 | -8.9 |
| Exp | 9.217 | 10.43 | 6.101 | 7.407 | 7.293 | 8.411 | 9.641 | 10.74 |

To assess the sensitivity of our results to the choice of the number of correlated electrons, we performed additional test calculations for Hg using a larger number. Table 1 shows that with 34 correlated electrons in occupied sites ($5s^2\ 5p^6\ 4f^{14}\ 5d^{10}\ 6s^2$) disponible to move into virtual orbitals, determined using the DIRAC default energy bounds, we got the value of 10.34 e.V. for the relativistic corrections using CC-DC. Now, considering 52 correlated electrons ($4s^2\ 4p^6\ 4d^{10}\ 5s^2\ 5p^6\ 4f^{14}\ 5d^{10}\ 6s^2$ ) while keeping the virtual orbitals intact, we got the value 10.35 e.V., which amounts to around 0.1% change. Thus, the original setting provides a balanced compromise between computational cost and accuracy, and the results are robust with respect to variations in the number of correlated electrons.

## 3.2 Correlation Effects in Relativistic and Non-Relativistic Frameworks

Table 2 reorganizes the data from Table 1 to highlight correlation effects in the relativistic and non-relativistic atomic systems.

Correlation effects generally increase ionization energies for Au, Hg, Tl, Bi, At, and Rn in both relativistic and non-relativistic frameworks. However, for Pb and Po, the behavior is dependent whether or not relativity was taken into account: correlation increases the ionization energy in the relativistic case but decreases it in the non-relativistic case. This element-specific behavior in the non-relativistic case demonstrates that correlation effects are not uniform across the periodic table and that their magnitude and sign depend also on the specific electronic structure of the atom.

When comparing with experimental values, correlation effects, as it was the case for relativistic effects in Table 1, improve predictions for atoms ending in s and $p_{1/2}$ orbitals in both relativistic and non-relativistic calculations (except for non-relativistic Pb). For atoms ending in $p_{3/2}$ orbitals, the pattern is less regular. However, for Bi, Po, At and Rn, correlation improves predictions in the relativistic case but worsens them in the non-relativistic case. This asymmetry demonstrates that correlation effects are not independent of relativistic effects: the same physical effect (electron correlation) produces different results depending on whether relativistic effects are included..

Table 2 : Correlation and non-correlation values of ionization energies(eV) of relativistic and non-relativistic atomic systems $_zX$ (where Z is the atomic number ranging from 79(Au) to 86(Rn)). Values in brackets denote the discrepancy with experimental values[13]. Δ represents the corresponding relative change.

|  | $_{79}$Au | $_{80}$Hg | $_{81}$Tl | $_{82}$Pb | $_{83}$Bi | $_{84}$Po | $_{85}$At | $_{86}$Rn |
|---|---|---|---|---|---|---|---|---|
| HF-NR | 5.915 | 6.829 | 5.068 | 6.208 | 7.358 | 8.532 | 9.738 | 10.98 |
|  | (35.8%) | (34.5%) | (16.9%) | (16.2%) | (0.9%) | (1.4%) | (1.0%) | (2.2%) |
| CC-NR | 7.019 | 8.326 | 5.404 | 6.016 | 8.660 | 8.084 | 9.758 | 11.47 |
|  | (23.8%) | (20.2%) | (11.4%) | (18.8%) | (18.7%) | (3.89%) | (1.2%) | (6.8%) |
| (CC - HF)$_{NR}$ | 1.104 | 1.497 | 0.336 | -0.192 | 1.302 | -0.448 | 0.020 | 0.490 |
| Δ% | 18.66 | 21.92 | 6.63 | -3.09 | 17.7 | -5.25 | 0.21 | 4.46 |
| HF-DC | 7.692 | 8.550 | 5.612 | 6.521 | 6.470 | 7.564 | 8.679 | 9.820 |
|  | (16.5%) | (18.0%) | (8.0%) | (12.0%) | (11.3%) | (10.1%) | (9.9%) | (8.6%) |
| CC-DC | 9.156 | 10.34 | 5.926 | 7.226 | 7.066 | 8.147 | 9.027 | 10.45 |
|  | (0.7%) | (0.9%) | (2.9%) | (2.4%) | (3.1%) | (3.1%) | (6.4%) | (2.7%) |
| (CC - HF)$_{DC}$ | 1.464 | 1.790 | 0.314 | 0.705 | 0.596 | 0.583 | 0.345 | 0.633 |
| Δ% | 19.03 | 20.94 | 5.60 | 10.81 | 9.95 | 7.71 | 3.97 | 6.45 |
| Exp | 9.217 | 10.43 | 6.101 | 7.407 | 7.293 | 8.411 | 9.641 | 10.74 |

We see that completing the s ($p_{1/2}$) orbital leads to an increase in the correlation corrections, as can be seen comparing (CC-HF)$_{DC}$ in Au to Hg (Tl to Pb) where the value increased from 1.464

(0.314) eV to 1.790 (0.705) eV.

## 3.3 Relative Importance of Relativistic and Correlation Effects

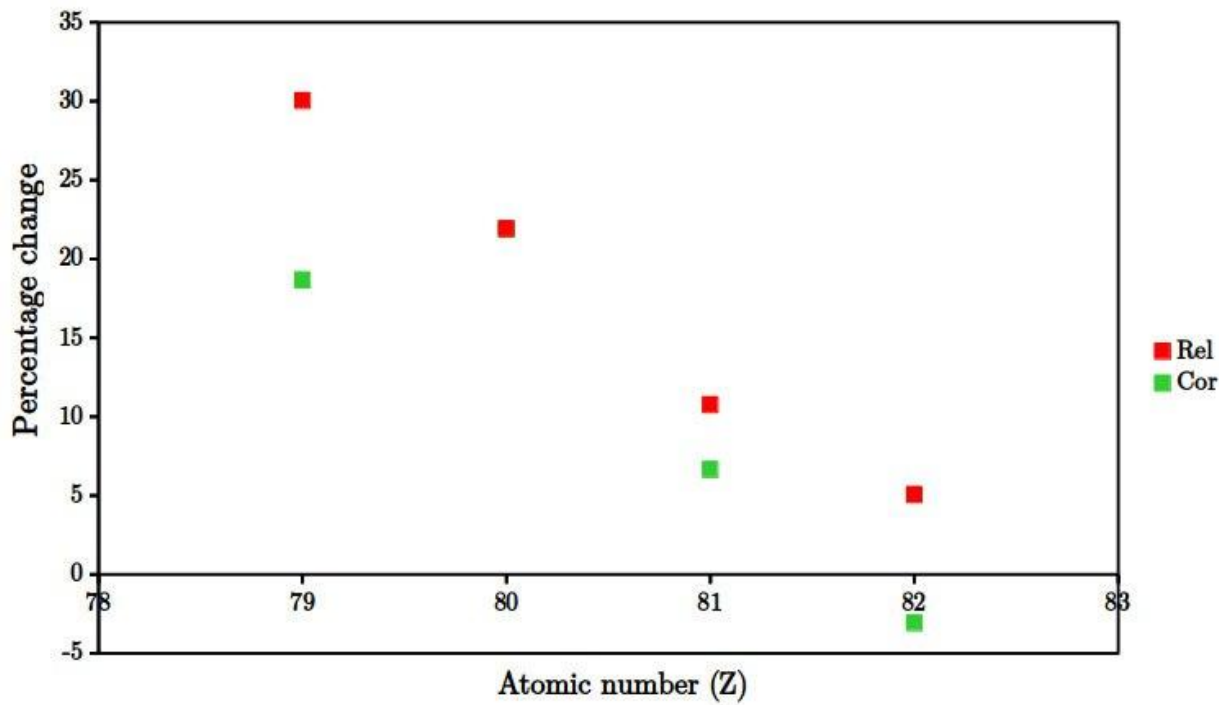

Figure 1: Percentages of Relativistic and Correlation effects on ionization energies versus the atomic number

In Fig.1, we contrast the relative "pure" relativistic effect in the absence of correlation: Δ(DC − NR)$_{HF}$ % (red), from Table 1, to the relative "pure" correlation effect in the absence of relativity: Δ (CC - HF)$_{NR}$% (green), from Table 2, restricting the data to ($_{79}$Au, $_{80}$Hg, $_{81}$Tl and $_{82}$Pb) where both relativistic and correlation corrections lead to an improvement approaching the experimental values.

We see that relativistic corrections are more important than correlation ones in Au, Tl and Pb, whereas both effects are equally significant for Hg. In addition, the two effects enhance each other in Au, Hg and Tl, whereas they partially cancel out in Pb. We see that relativity and correlation effects for atoms ending in s-orbitals (Z=79, 80) are larger than for atoms ending in p$_{1/2}$ -orbitals (Z=81,82). This is partially in line with the known fact that relativistic contraction of s-orbitals is larger than p$_{1/2}$-orbitals' due to how close the electron can approach the nucleus [15].

## 3.4 Non-Additivity of Relativistic and Correlation Effects

The central finding of this work concerns the non-additivity of relativistic and correlation effects. To investigate this, we compared three approaches: (1) the "combined" relativistic and correlation calculation (CC-DC), which includes both effects simultaneously; (2) adding the "pure correlation" correction in the absence of relativity ((CC - HF)$_{NR}$), from Table 2, to the relativistic uncorrelated (HF-DC) result, which treats correlation as an additive correction to the relativistic HF; and (3) adding the "pure relativistic" correction in the absence of correlation ((DC − NR)$_{HF}$), from Table 1, to the correlated non-relativistic (CC-NR) result, which treats relativity as an additive correction to the non-relativistic CC. If these effects were additive (i.e., independent), all three approaches would yield identical results.

Our results clearly demonstrate that this is not the case. Figure 2 shows the relative deviations from the experimental values corresponding to the three approaches described above (colored in red, black and blue accordingly).

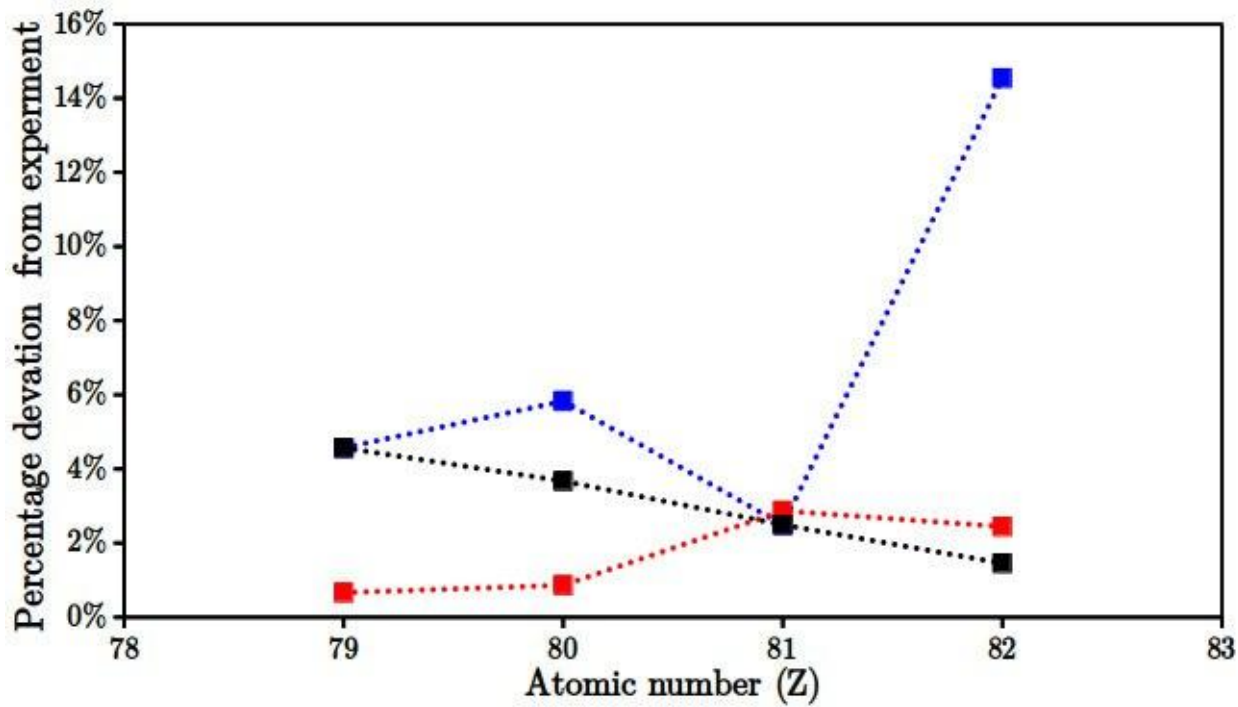

Figure 2: Relative deviations from the experimental values, (vertical axis), of the "combined" CC-DC calculations (red), and of the "total" corrections (black), computed by adding the "pure" non-relativistic correlation corrections to the relativistic HF result, and of the other "total" corrections (blue), computed by adding the "pure" uncorrelated relativistic corrections to the correlated CC-NR result.

We see that for $_{79}$Au and $_{80}$Hg, the combined CC-DC calculation (red points) yields better agreement with experiment than either additive approach (black/blue points). Conversely, for $_{81}$Tl, adding the individual corrections (black or blue) produces better results than the combined calculation, albeit the three approaches give approximately equal values. For $_{82}$Pb, approach 2 (black) gives the best result, whereas approach 3 is the worst. The discrepancy between the black and blue points in Fig. 2 reveals that the order in which corrections are applied, obtaining "total" corrections, does matter—the problem is non-commutative.

Furthermore, the magnitude of the discrepancy between the additive approaches and the combined calculation varies significantly across the periodic table, indicating that the degree of non-additivity depends on the atomic system.

This non-additivity reflects the fundamental non-linearity of the quantum many-body problem. When both relativistic and correlation effects are present, they interact in ways that cannot be predicted by simply summing their individual contributions. The physical reason for this non-additivity is that both relativistic and correlation effects modify the electronic structure and orbital energies. When both are present, the electronic structure is different from the sum of the structures obtained by applying each effect separately. This finding is consistent with earlier observations [8] that relativistic and correlation contributions are not independent, and our systematic study provides quantitative evidence of this interdependence across a series of heavy atoms.

### 3.5 Comparison with Previous Theoretical Data

Table 3 summarizes previously published theoretical ionization energies for the studied atoms, obtained with various methods where the values within brackets denote the deviations from the experimental values.

Table 3 : Other theoretical values of IE for the studied atoms, computed either dropping correlation effects with(HF-R)/without(HF-NR) relativistic effects using Hartree-Fock method, or  including correlation effects with(R)/without(NR) relativistic effects using Post-HF (essentially CC) or DFT. The values between brackets denote the deviations from experimental values.

|  | $_{79}$Au[6] | $_{80}$Hg[16] | $_{81}$Tl[17] | $_{82}$Pb[18] | $_{83}$Bi[19] | $_{84}$Po[19] | $_{85}$At[19] | $_{86}$Rn[20] |
|---|---|---|---|---|---|---|---|---|
| HF-NR | 5.953 (35.4%) | 6.883 (34.0%) | - | 6.666 (10.0%) | - | - | - | - |
| HF-R | 7.661 (16.9%) | 8.576 (17.8%) | 5.592 (8.3%) | 6.692 (9.7%) | - | - | 8.30 (13.9%) | - |
| (R - NR)$_{HF}$ | 1.708 | 1.693 | - | 0.026 | - | - | -1.66 | - |
| (Post-HF)$_{NR}$/(DFT)$_{NR}$ | 6.893 (24.2%) | 8.326 (20.2%) | - | - | - | - | - | - |
| (Post-HF)$_{R}$/(DFT)$_{R}$ | 8.930 (3.1%) | 10.29 (1.3%) | 6.109 (0.1%) | 6.971 (5.9%) | 7.266 (0.4%) | 8.427 (0.2%) | 9.308 (3.5%) | 10.56 (1.7%) |
| (R - NR)$_{Post-HF/DFT}$ | 2.037 | 1.964 | - | - | - | - | - | - |

Our CC-NR result for Hg (8.326 eV) matches the (Post-HF)$_{nr}$ value from the literature, which is expected since the Post-HF used in [16] was CCSD(T), with the same Hamiltonian, confirming consistency in non-relativistic correlated calculations. For Au, our CC-NR value (7.019 eV) differs slightly from the literature value (6.983 eV), likely due to a different Post-HF method (QCISD(T)) used in [6]. The close agreement between our HF-NR results for Bi–Rn and the (Post-HF)$_{r}$/(DFT)$_{r}$ literature values suggests that for these $p_{3/2}$-shell atoms, inclusion of relativity in post-HF/DFT calculations partly compensates for the neglect of correlation in our HF-NR treatment, hinting at fortuitous compensation. Actually, tables (1) and (2) show that for these atoms (Bi-Rn), relativistic corrections and electronic correlations work in opposite directions, in that relativity tends to decrease the ionization potential contrary to correlation which tends to increase it, which somehow justifies the almost complete canceling of errors.

The many gaps in Table 3 underline the scarcity of systematic studies on this topic, especially for Bi–Rn.

## 4. Conclusions

This comparative study of relativistic and correlation effects on ionization energies in heavy atoms reveals several important conclusions that have implications for computational chemistry and physics:

First, relativistic effects show consistent behavior across the periodic table: they increase ionization energies for lighter elements in this series (Au–Pb) due to orbital contraction and stabilization, and decrease them for heavier elements (Bi–Rn) due to spin-orbit coupling effects. This pattern holds regardless of whether correlation is included, though the magnitude of the effect differs between correlated and uncorrelated calculations.

Second, correlation effects show more complex behavior. Unlike relativistic effects, the sign of correlation corrections is not uniform across the periodic table. Moreover, correlation corrections can change sign when relativity is included, indicating that these effects are not independent. This demonstrates that one cannot simply predict the effect of correlation by examining non-relativistic calculations alone.

Third, and most importantly, relativistic and correlation effects are not additive. The combined effect of both corrections does not equal the sum of their individual contributions. This non-additivity indicates that these quantum effects interact in complex ways that depend on the specific atomic system. For most atoms in this study, computing the combined corrections simultaneously using the CC-DC method provides the best agreement with experiment. However, for specific cases (Z = 81 and 82, corresponding to Tl and Pb), the additive approach, treating correlation as an additive correction to the relativistic HF, yields superior results with far less computational effort compared to the combined effects strategy, suggesting that the optimal computational strategy may be system-dependent.

These findings underscore the importance of treating relativistic and correlation effects on equal footing in computational studies of heavy-element systems. Simple additive approaches may not capture the full complexity of these quantum effects, and careful validation against experimental data is essential. Our work emphasizes the non-additivity issue as a quantitative demonstration of the inherent non-linearity of the many-electron problem and provides a systematic numerical validation of this principle across the 6th-row elements

For accurate predictions of ionization energies and other properties in heavy elements, we recommend using methods that treat both effects simultaneously, such as relativistic coupled cluster theory. Future work should extend this analysis to other properties beyond ionization energy (such as electron affinities, excitation energies, and reaction barriers) and to other classes of heavy-element systems (such as transition metals and lanthanides) to determine the generality of these conclusions.

## Funding
This research did not receive any external funding.

## Data Availability

The data supporting the findings of this study are available within the article and from the corresponding authors upon request.

## Conflicts of Interest

The authors declare no conflict of interest.

## Acknowledgments

N.C. acknowledges support from the CAS PIFI fellowship and the Alexander von Humboldt Foundation.